# Examination of Boltzmann's H-Function: Dimensionality and Interaction Sensitivity Dependence, and a comment on his H-Theorem

**Shubham Kumar, Subhajit Acharya and Biman Bagchi***

**Solid State and Structural Chemistry Unit, Indian Institute of Science, Bengaluru-560012, Karnataka, India**
***Email: profbiman@gmail.com; bbagchi@iisc.ac.in**

**Boltzmann's H-Theorem, formulated 150 years ago in terms of H-function that also bears his name, is one of the most celebrated theorems of science and paved the way for the development of nonequilibrium statistical mechanics. Nevertheless, quantitative studies of the H-function, denoted by H(t), in realistic systems are relatively scarce because of the difficulty of obtaining the time-dependent momentum distribution analytically. Also, the earlier attempts proceeded through the solution of Boltzmann's kinetic equation, which was hard. Here we investigate, by direct molecular dynamics simulations and analytic theory, the time dependence of H(t). We probe the sensitivity of nonequilibrium relaxation to interaction potential and dimensionality by using the H-function H(t). We evaluate H(t) for three different potentials in all three dimensions and find that it exhibits surprisingly strong sensitivity to these factors. The relaxation of H(t) is long in 1D, but short in 3D. We obtain, for the first time, *a closed-form analytic expression for H(t) using the solution of the Fokker-Planck equation for the velocity space probability distribution and compare its predictions with the simulation results*. Interestingly, H(t) is found to exhibit linear response when vastly different initial nonequilibrium conditions are employed. The oft-quoted relation of H-function with Clausius's entropy theorem is discussed.**

One hundred and fifty years ago, in 1872, Boltzmann introduced his H-function and H-theorem, which heralded the birth of statistical mechanics. [1-3] It attempted to address many of the fundamental questions about the time evolution of nonequilibrium distribution functions and was successful in answering some of them. Since Newton's equations are time-reversible, the evolution of an initial nonequilibrium state irreversibly to a unique equilibrium state has always been a subject of intense discussions that involve, among other theories, Poincare's recurrence and Boltzmann's H-theorem. The former is resolved by noting the extraordinarily long time for the recurrence in a macroscopic system; the latter continues to be subjects of interest and lively speculation. Boltzmann's H-theorem has retained the status of one of the best well-known theorems of statistical mechanics, even though it is fairly restrictive and, in principle, strictly applicable only to a dilute gas. [1–3] However, H-function is general, although has not been evaluated in many applications.

In its original form, Boltzmann's theorem considers the nonequilibrium function H(t) defined by the following integration, [1–5]

$$\text{H}(t) = -\int d\mathbf{p}\, f(\mathbf{p},t) \ln f(\mathbf{p},t) \qquad (1)$$

where $f(\mathbf{p},t)$ is the time-dependent single-particle momentum (p) distribution function defined in the usual fashion by, $f(\mathbf{p},t) = \sum_{i=1}^{N} \delta(\mathbf{p}-\mathbf{p}_i)$. It is a singlet distribution function.

Note the simplicity of Eq. (1): it does not contain integration over position coordinates. Thus, it is not an average over the full phase space. If we need to make any connection with thermodynamics like entropy, then this point becomes useful, as we discuss later. Thus, the nonequilibrium state is nonequilibrium with respect to momentum only. In essence, the H-function assumes either the system is homogeneous or the position relaxation is much slower than momentum relaxation.

Boltzmann's H-Theorem states that if the distribution $f(\mathbf{p},t)$ satisfies Boltzmann's transport equation, then [1–8]

$$\frac{dH}{dt} \geq 0 \qquad (2)$$

The equality sign is satisfied only at equilibrium when the distribution attains the Maxwell form. That is, when $f(\mathbf{p},t)$ is not an equilibrium distribution, the function H will continuously increase till the equilibrium distribution is reached. Thus, the time dependence of H(t) can be a quantitative measure of the rate of approach of $f(p,t)$ to equilibrium. Differentiation of the H function is given by,

$$\frac{dH(t)}{dt} = -\int d^3\mathbf{p} \frac{\partial f(\mathbf{p},t)}{\partial t}[1+\ln f(\mathbf{p},t)] \tag{3}$$

Since, at equilibrium, $\partial f/\partial t = 0$ implies $\frac{dH}{dt} = 0$ at equilibrium.

In the following, we briefly outline a proof of the H-theorem. We start with the Boltzmann kinetic (or transport) equation that, in its final form, is written as,

$$\left[\frac{\partial}{\partial t} + \frac{\mathbf{p}}{m}\cdot\nabla_r + \mathbf{F}\cdot\nabla_p\right]f^{(1)}(\underline{r},\underline{p},t) = \left(\frac{\partial f^{(1)}}{\partial t}\right)_{coll}, \tag{4}$$

where the collisional term is written as,

$$\left(\frac{\partial f^{(1)}}{\partial t}\right)_{coll} = \iiint d\mathbf{p}_2 d\mathbf{p}'_1 d\mathbf{p}'_2 \delta(\mathbf{P}-\mathbf{P}')\delta(E-E')|T_{fi}|^2 \left(f^{(2)}_{1'2'} - f^{(2)}_{12}\right) \tag{5}$$

One can easily identify the energy and momentum constraints. Note that otherwise, we use the indices for the initial state and the final states, as "$i$" and "$f$". The T-matrix gives the transition probability. At this stage, we employ the assumption of molecular chaos to express the two-particle distribution function $f^{(2)}$ in terms of the product of two one-particle distribution functions. By certain algebraic manipulations, we are led to the following condition for the time derivative of the H function, [4]

$$\frac{dH}{dt} = -\frac{1}{4}\int d^3\mathbf{p}_2 \, d^3\mathbf{p}'_1 \, d^3\mathbf{p}'_2 \delta(\mathbf{P}_f - \mathbf{P}_i)\delta(E_f - E_i)|T_{fi}|^2 \left(f^{(1)}_{2'}f^{(1)}_{1'} - f^{(1)}_{2}f^{(1)}_{1}\right) \times \left[\ln\left(f^{(1)}_{2}f^{(1)}_{1}\right) - \ln\left(f^{(1)}_{2'}f^{(1)}_{1'}\right)\right]$$

$$\tag{6}$$

The integrand in Eq. (6) is never positive, so the H-theorem is satisfied. Thus, in the above derivation, H-theorem is tied to the validity of the Boltzmann kinetic equation.

*Despite the formidable character of Eq. (6), a calculation of the H-function itself is easy to implement, and its quantitative evaluation can be carried out without any approximation.* We first need to obtain or define a nonequilibrium time-dependent momentum distribution function. The simplicity lies in the singlet character of the density distribution. Note that the criticism of molecular chaos assumption that is explicit in Eq. (6) is not at all required in the definition of H-function given in Eq. (1). Thus, while it is very difficult to evaluate Eq. (6), Eq. (1) can be evaluated by computer simulations by creating various nonequilibrium momentum distribution functions. As the time-dependent nonequilibrium distribution approaches the Maxwell-Boltzmann velocity distribution, the H-function also evolves simultaneously and provides a measure of the nature and time scale of the relaxation. Up to date, we are aware of only a few explicit evaluations of this function. [4,9–12] This is because initial studies attempted to obtain H(t) through the solution of Boltzmann kinetic equation, which is hard. [4] There have also been several studies using a generalized Boltzmann H-function defined differently where the following expression has been used, [5,13,14]

$$H_{GB}(t) = -\int dx P(x,t) \ln\left(\frac{P(x,t)}{P_{eq}(x)]}\right). \tag{7}$$

This generalized form of the Boltzmann's H-function serves a similar purpose as the original Boltzmann's H-function. Here, $P_{eq}(x)$ is the equilibrium distribution of a given variable *x*. The variable "*x*" has often been assumed to be a position variable or a combination thereof. This is the form advocated in the well-known monograph of Kubo and Toda. [5]

In this work, we shall be concerned with quantitative aspects of nonequilibrium velocity relaxation using the *original H-function*. We ask the following questions: (i) What are the time scales of the growth of this function? We imagine that this would be related to the friction or diffusion constant of the gas, but the quantitative dependence is not clear. (ii) One, fortunately, knows the exact solution of the momentum space Fokker-Planck equation which, even in such a simple case, gives a non-trivial time dependence of the single-particle momentum-space distribution function, $f(p,t)$. We would like to check the reliability of this description. (iii) What is the range of validity of Boltzmann's H-theorem for interacting systems? We ask this question because Boltzmann's entire treatment was for dilute gas, and the H-function does not contain any spatial variables. Furthermore, Boltzmann's proof used his kinetic equation, which we know has limited validity. (iv) Does H(t) satisfy linear response? This could be an intriguing question because Maxwell velocity distribution can be regarded as stabilized by a harmonic force constant $\frac{k_B T}{m}$. (v) How can we relate this to entropy in a rigorous way, given that the original H-function contains only velocity?

In order to understand, examine and employ the H-Theorem, it is essential to create a proper nonequilibrium momentum distribution function in an isolated system. We studied several such distribution functions and evaluated H(t) for different 3D, 2D, and 1D systems, namely (i) Lennard-Jones, (ii) soft-sphere, and (iii) hard-sphere corresponding to initial nonequilibrium velocity distributions. The simulation details have been described in Supplementary Material.

In **Figures 1 (a) to 1(c)**, we show the evolution of H(t) for different systems of dilute gases (at reduced density, $\rho^* = 0.10$ and average reduced temperature, $T^*=2.0$, which is obtained by a procedure described below) where not only the interaction potential is varied from system to system but also the dimensionality of the systems is changed. The initial nonequilibrium state is created by taking the amplitude of the velocities of all the particles exactly the same; the magnitude is in accordance with the equipartition theorem and temperature. This allows us to carry out the simulation in the microcanonical (NVE) ensemble. We see that in all the cases, the H-function sharply increases in a short time and subsequently monotonically approaches the equilibrium value (shown by a black dotted line) at a longer time, which is the equilibrium value at the chosen temperature. Note that the time scales have been converted from reduced unit to real unit (in ps) using the argon parameters. In **Figures 1 (d) to 1(f)**, we show normalized H(t), defined as $\left((H(t)-H_{eq})/(H(0)-H_{eq})\right)$ and its exponential fit using the expression $A_1 + A_2 \exp(-t/\tau)$. The fitting parameters are given in Supplementary Material.

It is evident from Figure 1 that the nonequilibrium relaxation function H(t) is sensitive to both potential and dimensionality. For 3D systems, the approach of H(t) to the equilibrium value is the fastest, while for 1D systems, it is the slowest. We also find that in all cases, the relaxation times of the systems interacting via Lennard-Jones potential are faster than those interacting via hard-sphere potential. This slow relaxation of H(t) for hard-sphere systems deserves further studies.

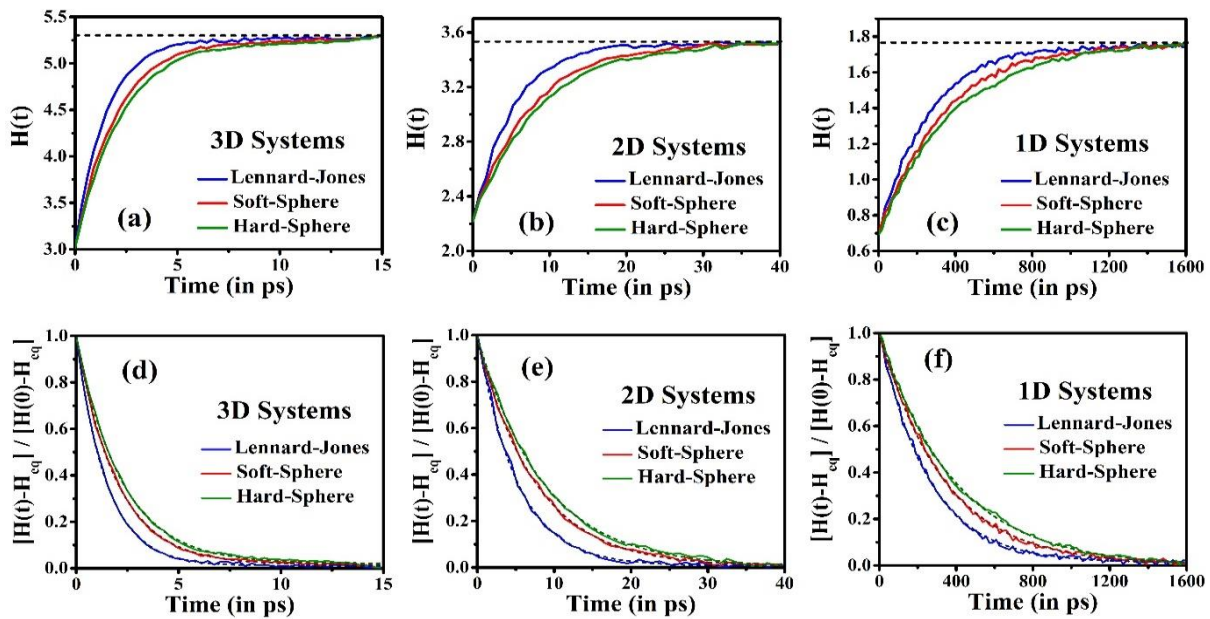

**Figure 1. Time evolution of H(t) obtained via computer simulations for (a) 3D, (b) 2D, and (c) 1D systems of dilute gases (at reduced density $\rho^* = 0.10$, and average reduced temperature, T\*=2.0) interacting with Lennard-Jones, soft-sphere, and hard-sphere potentials. In all the cases, the H-function increases monotonically and then attains equilibrium at a longer time, which is the equilibrium value at the final temperature, shown by black dashed lines in the figures. (d)-(f) Time evolution of the normalized H(t) for the systems shown in (a)-(c). Dotted lines show the exponential fitting of normalized H(t). Note the different time scales in each case.**

While the validity of Boltzmann's theorem as the arrow of time was never in doubt at very low-density gas, quantitative estimates of time scales were not known. This in itself is an interesting issue because as density decreases, collisions become rare, and the rate of approach to equilibrium becomes slow. But the nature and time scales at higher densities remain unexamined. In order to examine such issues, we have performed simulations with two different initial nonequilibrium momentum distributions (as shown in **Figures 2(a) and 2(b)**). In **Figure 2(c),** we show the temporal evolution of H(t) for 1D Lennard-Jones systems with two different initial nonequilibrium conditions. While it seems H(t) shows distinct features in the two cases, the normalized H(t) exhibits similar behavior (as shown in **Figure 2(d)**) with almost same relaxation times (the relaxation times are given in Supplementary Material). Besides it, we have checked the validity of the linear response of H(t) in 2D and 3D systems. Thus, it is fair to say that H(t) exhibits a linear response.

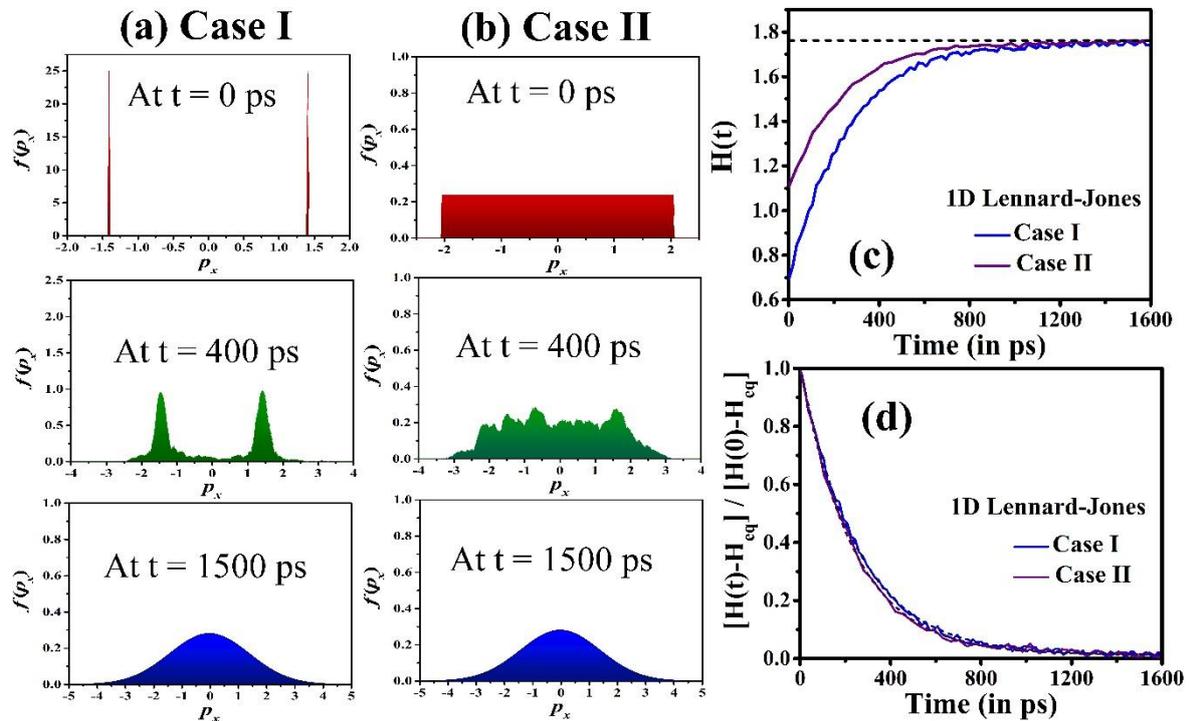

**Figure 2. Evolution of the momentum distribution at different times when initially all particles have (a) same amplitude of velocity, and (b) uniform distribution of velocity. (c) Time evolution of H-function for 1D Lennard-Jones systems with two different initial distributions. (d) Time evolution of the normalized H-function, showing the validity of linear-response theory. Dotted lines show the exponential fitting of normalized H(t). The final distribution in each case is Maxwellian and identical.**

The H function and entropy have been associated with each other from the beginning because both define a direction of time. Clausius's statement asserts that the entropy of an isolated system can only increase with time. In fact, for a one-component ideal gas, one can derive a simple relation between the two. For a 3D ideal gas at equilibrium, the velocity distribution is Maxwellian, and one can easily evaluate the H function at equilibrium to obtain,

$$H_{eq} = -\ln\left(\frac{1}{2\pi m k_B T}\right)^{3/2} + \frac{3}{2}. \tag{8}$$

We can obtain the entropy per particle of an ideal gas from the Sackur-Tetrode equation,

$$S^{id} = k_B \ln\left(2\pi m k_B T\right)^{3/2} + k_B \ln V + \frac{5}{2} \tag{9}$$

Thus, we obtain the following relation,

$$S^{id} = k_B H_{eq} + \text{constant}, \tag{10}$$

at constant volume V. We emphasize that this remarkable relation holds only for an ideal gas at equilibrium. There is an "ln V" term absorbed in the "constant" factor of Eq.(10).

As both the functions increase as an initial nonequilibrium state evolves to essentially the same values in the equilibrium state, it is natural to look for a relationship between H and S, even as a function of time. However, there has been no convincing proof that such a relation should indeed exist. The only exact statement we can make is that both can serve as the arrow of time. A strictly valid definition for time-dependent entropy is not available. One can attempt to define evolving entropy of a sub-system that is in contact with a bath that is governed by faster dynamics. For example, we can change the temperature of the system in a controlled manner, slowly, such that one can define entropy in the intermediate states. However, that remains problematic because H(t), on the other hand, is defined for an isolated system.

We first consider the case when the value of H(t) is only slightly different than the equilibrium value, i.e., when $f(p,t)$ is close to Maxwell distribution. Let us define $\delta f(p,t) = f(p,t) - f_M(p)$. The above analysis suggests that we can attempt a stochastic approach. We now use the Fokker-Planck equation in the momentum space for *f(p,t)*. The equation is then is a Fokker-Planck equation [5,15]

$$\frac{\partial \delta f(p,t)}{\partial t} = \zeta \left( \frac{\partial}{\partial p} [\frac{p}{m} + <E> \frac{\partial}{\partial p}] \right) \delta f(p,t) \quad (11)$$

The above equation has the solution,

$$f(p,t) = \frac{1}{(2\pi m k_B T (1-\Gamma^2(t)))^{3/2}} \exp(-[p - p_0 \Gamma(t)]^2 / (2m k_B T (1-\Gamma^2(t)))) \quad (12)$$

where, $\Gamma(t) = e^{-\zeta t}$. Fortunately, we can obtain a closed-form analytic expression for H(t) using the Fokker-Planck equation for the momentum space probability distribution as follows (a detailed description is given in Supplementary Material),

$$H(t) = -\frac{3}{2} \ln \left( \frac{1}{2\pi m k_B T (1-\exp(-2\zeta t))} \right) + \frac{3}{2}. \quad (13)$$

We can see that this form already predicts the rapid rise of the H function at short times. Thus, the Fokker-Planck equation can capture the time dependence of H(t) through the distribution function *f(p,t)*. We calculated the value of friction $(\zeta)$ from the diffusion coefficient (*D*) using Einstein relation $D = k_B T / \zeta$ and put it in Eq. (13) to obtain H(t). **In Figure 3**, we compare the results obtained via simulation and Eq. (13) for 3D and 2D systems. In the case of 3D systems, we find that the Fokker-Planck equation provides a reasonable description for Lennard-Jones and soft-sphere systems but fails for hard-sphere systems. We further observe that for 2D and 1D systems, the Fokker-Planck equation-based description of H(t) fails. This failure needs further studies.

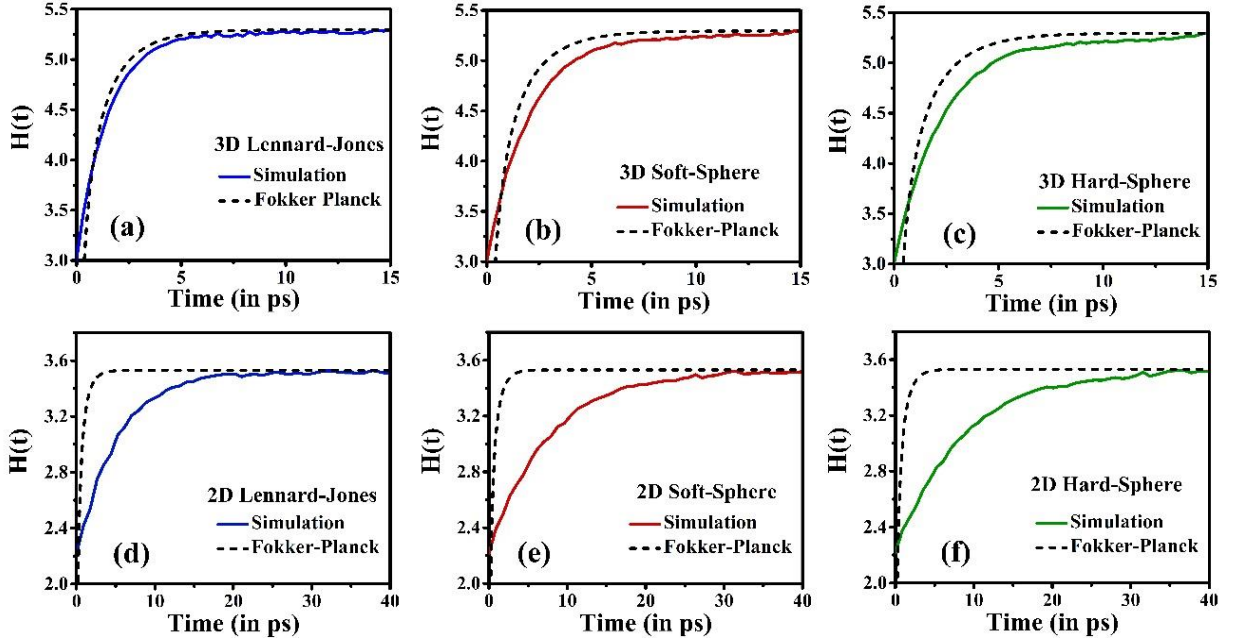

**Figure 3: Comparison of H(t) obtained via simulation and Fokker-Planck equation for 3D and 2D systems. We employ Eq. (13) to obtain the variation of H(t) with $\zeta$ obtained from simulation. Note that rapid evolution of H(t) in three dimensions while the slower decay in 2D. In 1D, the decay becomes much slower.**

The breakdown of the Fokker-Planck equation in one- and two-dimensional systems can be quite instructive. We attribute this failure to the emergence of memory effects even at low densities. Thus, the Markovian Fokker-Planck equation needs to be replaced by the non-Markovian equation. [16,17] The Markovian description used here gives rise to too large value of the friction at short times.

Let us again turn to the relationship of H(t) with entropy. By Boltzmann's formula, entropy is given by the total number of states. We can evaluate the entropy if we assume that the distribution changes infinitesimally slowly, say at $\zeta \sim 0$. In that limit, we can calculate the total number of states and hence the entropy. That is, we need to calculate the number of configurations $\Omega$ corresponding to a given slowly evolving momentum-space distribution function.

It is interesting to inquire about the range of validity of the H-Theorem. From its original proof by Boltzmann using his transport equation, it is not clear that the theorem remains valid at higher densities. We verified the validity at $\rho^* = 0.20$. However, note that the relation between the H-function and entropy may not hold at high densities because the total entropy is increasingly dominated by intermolecular correlations as the density is increased progressively.

The sensitivity of H(t) to the interaction potential, of course, reflects the sensitivity of the relaxation of *f(p,t)*. The difference between Lennard-Jones and hard-sphere systems has been examined earlier with the Enskog approximation. [18] This is an interesting aspect we believe deserves further examination. It represents at least partly the dependence on the range of potential. This agrees with the much slower relaxation in one dimension where the number of nearest neighbours is limited. Earlier studies in 1D have pointed out the anomalous nature of particle displacements in one dimension. [19,20]

As remarked earlier, there appears to be surprisingly few numerical studies of H(t). The present study fills this lacuna and serves to provide a detailed understanding of Boltzmann's H-theorem, which is one of the most celebrated theorems of science and paved the way for developing nonequilibrium statistical mechanics. We find in every case Boltzmann's H theorem is verified. We also find dimension and interaction potential dependent time scales which for a density of $\rho^* = 0.10$ ranges from a few picoseconds in three dimensions to several hundred picoseconds in one dimension. The sensitivity of H(t) on the potential and dimensionality of the system should help to understand many basic aspects of nonequilibrium phenomena. It would be fascinating to employ it in many areas, such as plasma physics and active matter. [12,21,22]Finally, the failure of Fokker-Plack description points to the importance of non-Markovian or memory effects, which surprisingly is influenced by the interaction potential.

**Acknowledgment:** BB thanks the Science and Engineering Research Board (SERB), India, for the National Science Chair Professorship and the Department of Science and Technology, India, for partial research funding. SK thanks the Council of Scientific and Industrial Research (CSIR), India, for research fellowship. SA thanks IISc for research fellowship.

# Supplementary Material

# Examination of Boltzmann's H-Function: Dimensionality and Interaction Sensitivity Dependence, and a comment on his H-Theorem

**Shubham Kumar, Subhajit Acharya and Biman Bagchi\***

**Solid State and Structural Chemistry Unit, Indian Institute of Science, Bengaluru-560012, Karnataka, India**
**\*Email: bbagchi@iisc.ac.in**

## S1. Simulation details

We have carried out a series of nonequilibrium molecular dynamic simulations of dilute gases in one, two, and three dimensions in order to study the evolution of H-function. Our model system consists of 10000 particles in each case. We have carried out these simulations in the microcanonical ensemble (constant N, V, and E) by applying the usual periodic boundary conditions. We choose three different radially symmetric potentials: (a) Lennard-Jones (LJ), (b) soft-sphere (SS), and (c) hard-sphere (HS) to define the interaction between any two particles. The potential forms are given below

for Lennard-Jones, $$U_{LJ}(r) = 4\varepsilon\left[\left(\frac{\sigma}{r}\right)^{12} - \left(\frac{\sigma}{r}\right)^{6}\right], \tag{S.1}$$

for soft-sphere, $$U_{SS}(r) = \varepsilon\left(\frac{\sigma}{r}\right)^{12}, \tag{S.2}$$

and, for hard-sphere, $$\begin{aligned} U_{HS}(r) &= \infty \quad if \ r < \sigma \\ &= 0 \quad if \ r > \sigma \end{aligned} \tag{S.3}$$

We have taken the diameter and mass of the particles equal to unity in the reduced unit i.e., $\sigma^* = 1.0$ and $m^* = 1.0$. For Lennard-Jones and soft-sphere potentials, we keep the interaction strength $\varepsilon^* = 1.0$. The reduced density $\left(\rho^* = \rho\sigma^{\dim}/m\right)$ is taken as 0.10 for all the systems, where superscript 'dim' represents the dimensionality of the system.

The initial nonequilibrium state is created by taking the amplitude of the velocities of all the particles exactly the same; the magnitude is in accordance with the equipartition theorem corresponding to the reduced temperature $T^* = k_B T/\varepsilon = 2.0$. For 3D and 2D systems, it has been illustrated in **Figure S1** while for the 1D system, it has been shown in Figure 2 (a). This approach allows us to carry out the simulation in the microcanonical (NVE) ensemble.

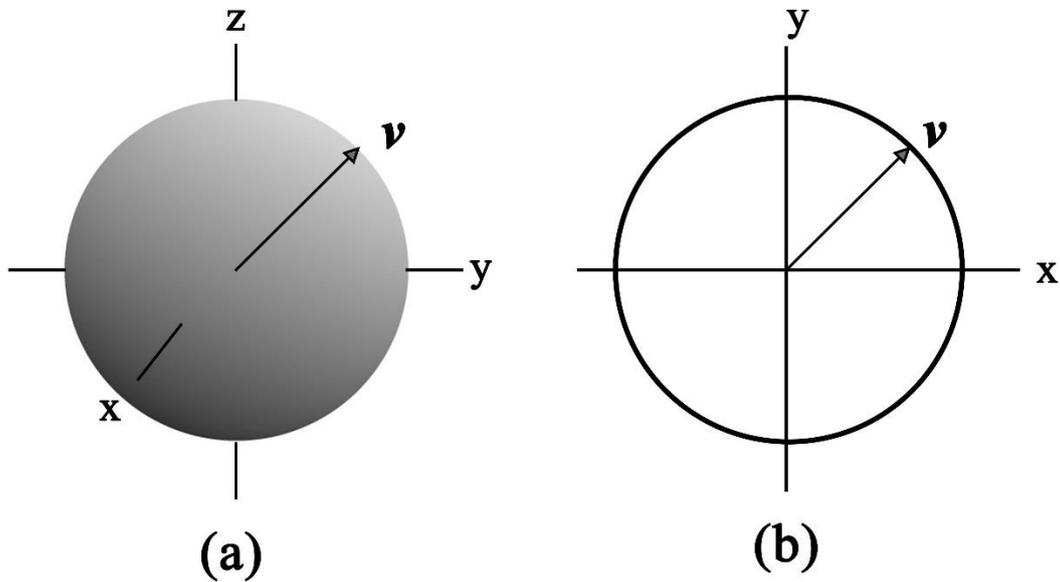

**Figure S1. An illustration of the initial nonequilibrium state created by taking the amplitude of the velocities of all the particles exactly the same for (a) 3D and (b) 2D systems. Thus, only the directions of the velocities of the individual particles are selected to the randomly distributed, on a sphere (for 3D) and on a circle (for 2D).**

Boltzmann's H-function defined as, $H(t) = -\int dp_x dp_y dp_z f(\{p\},t) \ln f(\{p\},t))$ is a relatively simple function; however, its evolution for 3D systems is quite complicated as it involves a three-dimensional integral. Besides it, the calculations become non-trivial because of the convergence problem, as we are studying the distribution functions. In order to get a converged result, we have to simulate large systems for a long time.

In this work, the time scales have been converted from reduced time (t*) to real-time (t) using relation $t^* = t\sqrt{\frac{\varepsilon}{m\sigma^2}}$, where the values of $\varepsilon, m$ and $\sigma$ have been taken corresponding to that of argon atom i.e., $\varepsilon/k_B = 119.8\ K, m = 0.03994\ kg/mol$ and $\sigma = 3.405\times 10^{-10} m.$

## S2. Fitting parameters of normalized H(t)

We show the normalized H(t) defined as $\left((H(t)-H_{eq})/(H(0)-H_{eq})\right)$ and its exponential fit using the expression $A_1 + A_2 \exp(-t/\tau)$ for 3D, 2D, and 1D systems in Figures 1(d) to 1(f). We provide the fitting parameters in **Table S1, S2, and S3** for 3D, 2D, and 1D systems, respectively. **Figure S2** shows the time evolution of the normalized H-function (in logarithmic scale). We find that the time evolution of the H-function in the case of Lennard-Jones potential is quite different from that of hard-sphere potential. We should also remark here that the approach of equilibrium in 1D is substantially slower than that in 2D and 3D.

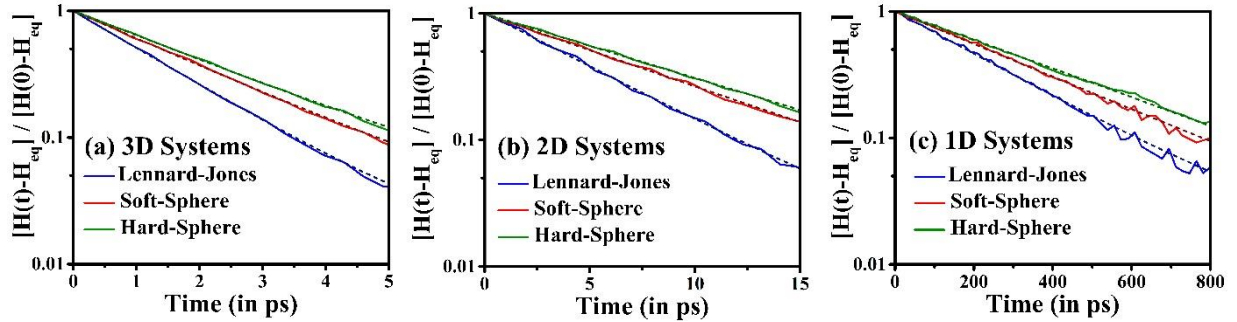

**Figure S2. Time evolution of the normalized H-function (in logarithmic scale) for (a) 3D, (b) 2D, and (c) 1D systems of dilute gases (at reduced density $\rho^* = 0.10$, and average reduced temperature, T\*=2.0) interacting with Lennard-Jones, soft-sphere, and hard-sphere potentials. Dotted lines show the exponential fitting to the normalized H(t).**

**Table S1. The fitting parameters for the normalized H(t) for three-dimensional systems.**

| Systems | $A_1$ | $A_2$ | τ (ps) |
|---|---|---|---|
| Lennard-Jones | 0.011 | 0.989 | 1.461 |
| Soft-sphere | 0.015 | 0.985 | 1.952 |
| Hard-Sphere | 0.021 | 0.979 | 2.181 |

**Table S2. The fitting parameters for the normalized H(t) for two-dimensional systems.**

| Systems | $A_1$ | $A_2$ | τ (ps) |
|---|---|---|---|
| Lennard-Jones | 0.013 | 0.987 | 5.064 |
| Soft-sphere | 0.017 | 0.983 | 7.512 |
| Hard-Sphere | 0.022 | 0.978 | 8.483 |

**Table S3. The fitting parameters for the normalized H(t) for one-dimensional systems.**

| Systems | $A_1$ | $A_2$ | τ (ps) |
|---|---|---|---|
| Lennard-Jones | 0.015 | 0.985 | 253.192 |
| Soft-sphere | 0.019 | 0.981 | 333.127 |
| Hard-Sphere | 0.024 | 0.976 | 386.723 |

## S3. Validity of linear-response theory

Similar to Figure 2, in **Figures S3 (a) and S3 (b)**, we show the time evolution of H(t) for 1D soft-sphere and 1D hard-sphere systems, respectively, with two different initial nonequilibrium conditions. In case I, we have taken the amplitude of the velocities of all the particles exactly the same. In contrast, in case II, we have taken a uniform rectangular velocity distribution to create the initial nonequilibrium state. In **Figures S3 (c) and S3 (d)**, we show the temporal evolution of normalized H(t) with two different initial nonequilibrium conditions and its exponential fit for 1D soft-sphere and hard-sphere systems. The relaxation times obtained by the exponential fit of normalized H(t) are given in **Table S4**.

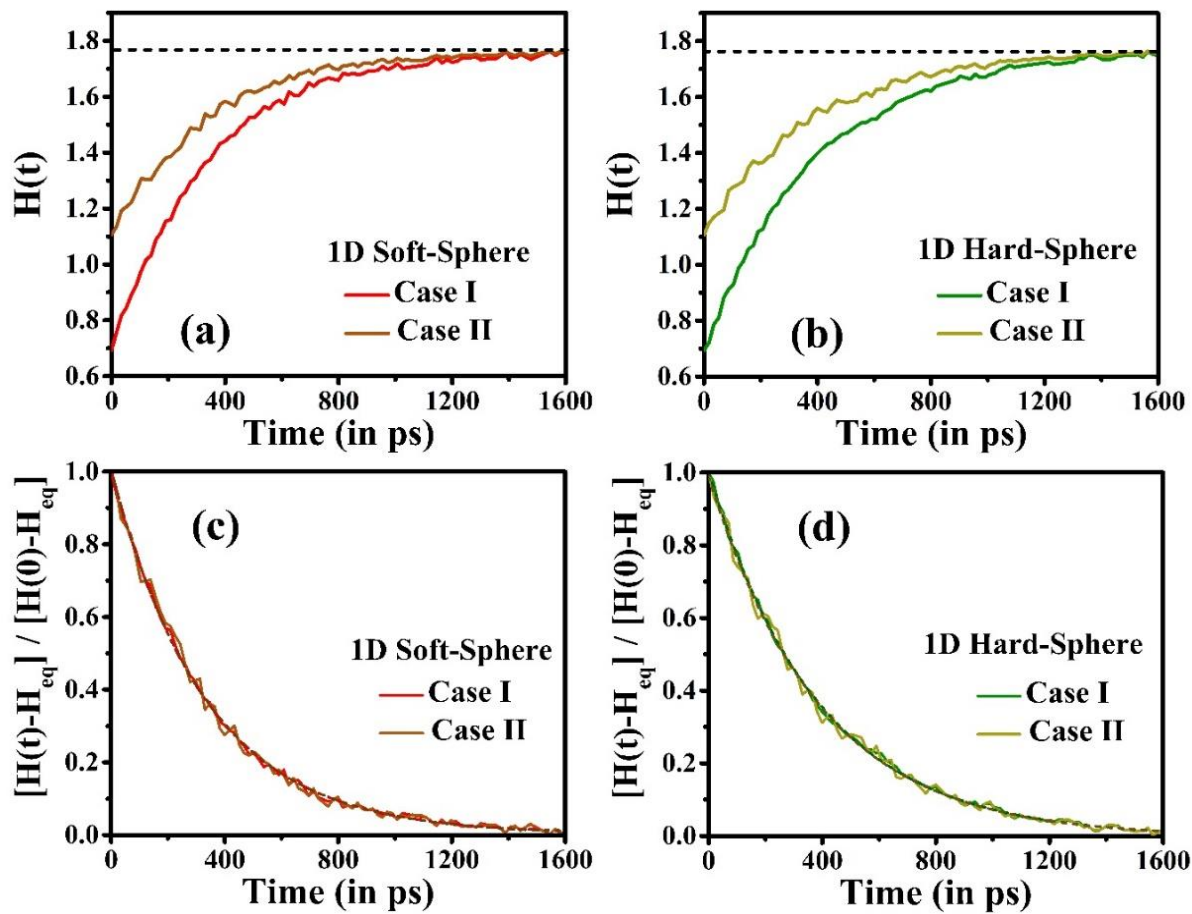

**Figure S3.** Time evolution of H-function for 1D (a) soft-sphere, and (b) hard-sphere systems with two different initial distributions. Time evolution of the normalized H-function, showing the validity of linear-response theory for (c) soft-sphere, and (d) hard-sphere systems. Dotted lines show the exponential fitting of normalized H(t).

**Table S4.** The relaxation time obtained by exponential fit of normalized H(t) for one-dimensional systems with two different initial nonequilibrium conditions.

|  | $\tau$ (ps) | |
| --- | --- | --- |
| Systems | Case I | Case II |
| Lennard-Jones | 253.192 | 249.125 |
| Soft-sphere | 333.127 | 334.725 |
| Hard-Sphere | 386.723 | 389.646 |

We have further checked the validity of linear response of H(t) for 2D and 3D systems. In **Figure S4**, we show the time evolution of H(t) for 2D and 3D Lennard-Jones systems having two different initial nonequilibrium momentum distributions: (i) all particles having the same amplitude of momentum with different directions, and (ii) uniform rectangular distribution of momentum. The temporal evolution of normalized H(t) shows the validity of linear response theory. Similar results have been obtained for soft-sphere and hard-sphere systems in 2D and 3D. In **Table S5**, we provide the relaxation times obtained by the exponential fit of normalized H(t) for 2D and 3D Lennard-Jones systems.

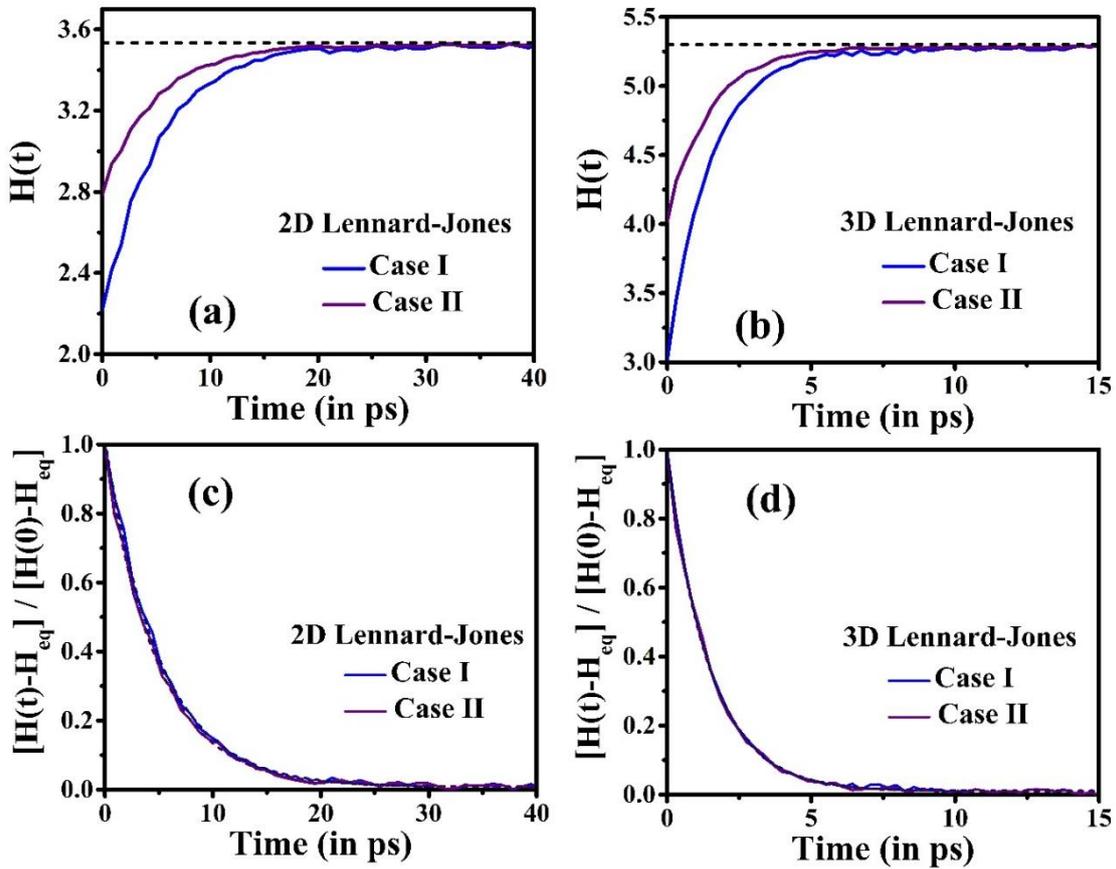

**Figure S4. Time evolution of H-function for (a) 2D, and (b) 3D Lennard-Jones systems with two different initial distributions. Time evolution of the normalized H-function, showing the validity of linear-response theory for (c)2D, and (d) 3D Lennard-Jones systems.**

**Table S5. The relaxation time obtained by exponential fit of normalized H(t) for 2D and 3D Lennard-Jones systems with two different initial nonequilibrium conditions.**

| Systems | τ (ps) | |
| --- | --- | --- |
| | Case I | Case II |
| 2D Lennard-Jones | 5.064 | 4.996 |
| 3D Lennard-Jones | 1.461 | 1.474 |

## S4. Density dependence of H(t)

According to Boltzmann's H-theorem, for a dilute gas, H(t) is an ever-increasing function of time till it reaches an equilibrium value. However, for dense fluids, the validity of the H-theorem has remained doubtful. In order to check the validity of Boltzmann's H-theorem for dense fluids, we have performed simulations at higher densities. **Figure S5** shows the temporal evolution of the H-function of 3D Lennard-Jones systems at three different densities, $\rho^* = 0.10, 0.20$ *and* $0.30$. We find that the H-theorem remains valid even at higher densities. Further, as the density increases, the momentum relaxation becomes faster.

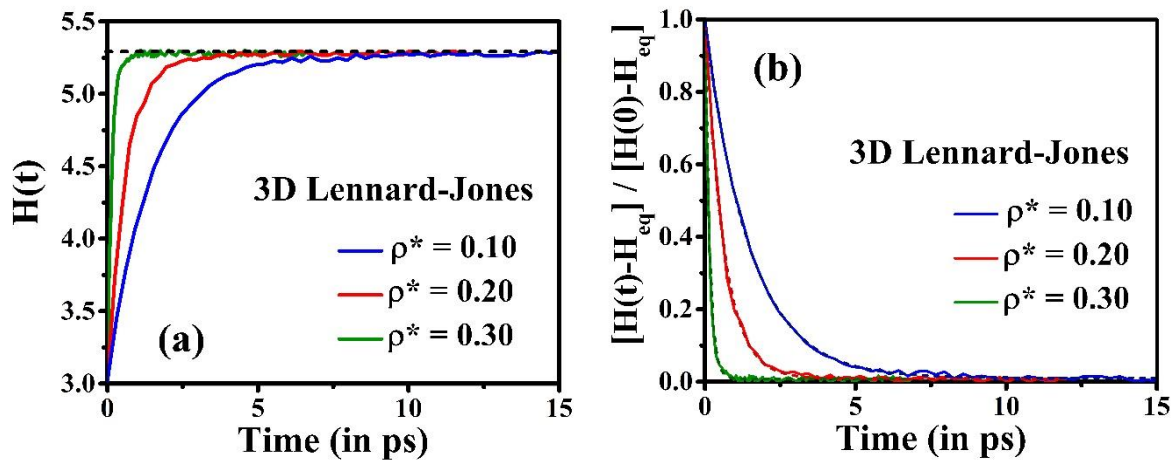

**Figure S5. (a) Variation of H(t) as a function of time for 3D Lennard-Jones systems at different densities, $\rho^* = 0.10, 0.20$ and $0.30$. (b) Time evolution of the normalized H(t) for the systems shown in (a). Dotted lines show the exponential fitting of normalized H(t).**

The fitting parameters for the exponential fitting of the normalized H(t) are provided in **Table S6.**

**Table S6. The fitting parameters for the normalized H(t) for three-dimensional Lennard-jones systems at different densities.**

| Density | $A_1$ | $A_2$ | $\tau$ (ps) |
|---|---|---|---|
| $\rho^* = 0.10$ | 0.013 | 0.987 | 1.461 |
| $\rho^* = 0.20$ | 0.011 | 0.989 | 0.615 |
| $\rho^* = 0.30$ | 0.016 | 0.984 | 0.145 |

## S5. Closed-form analytic expression for H(t) using the Fokker-Planck equation

In the absence of external potential, the solution of the Fokker-Planck equation in the momentum space provides the expression for the momentum distribution function as follows,

$$f(\{p\},t) = \left[\frac{1}{2\pi m k_B T(1-\exp(-2\zeta t))}\right]^{3/2} \exp\left\{-\frac{(p_x - p_{x0}\exp(-\zeta t))^2 + (p_y - p_{y0}\exp(-\zeta t))^2 + (p_z - p_{z0}\exp(-\zeta t))^2}{2 m k_B T(1-\exp(-2\zeta t))}\right\}$$

(S.4)

Here $m$ is the mass of the particle; $p_{x0}$, $p_{y0}$, $p_{z0}$ denote the x-component, y-component, and z-components of initial momentum, and $p_x$, $p_y$, $p_z$ are the current momentum of the particle at time t along x, y, and z directions. $T$ is the absolute temperature of the system and $\zeta$ is the friction experienced by the particle.

On the other hand, the Boltzmann H function in three dimensions is defined as,

$$H(t) = -\int dp_x dp_y dp_z f(\{p\},t) \ln f(\{p\},t) \qquad (S.5)$$

We next substitute Eq. (S.4) in Eq. (S.5) and perform an integration over $p_x$, $p_y$, and $p_z$ to obtain, after a lengthy algebraic manipulation, the following simplified equation

$$H(t) = -\frac{3}{2}\ln\left(\frac{1}{2\pi m k_B T(1-\exp(-2\zeta t))}\right) + \frac{3}{2}. \qquad (S.6)$$

It is interesting to note that Eq. (S.6) is independent of the initial velocities. Therefore, the major limitation of this approach is that it cannot capture the effects of the initial distributions used to generate a nonequilibrium configuration.

## S6. Normalized velocity autocorrelation function (VACF)

In **Figure S6**, we show the normalized velocity autocorrelation functions for 3D, 2D, and 1D Lennard-Jones systems. In the case of 1D system, the diffusion constant obtained via VACF is very low compared to that in the 3D system, leading to a very high value of friction obtained using the Einstein relation $D = k_B T/\zeta$. Thus, the H(t) obtained by the closed-form analytical solution of the Fokker-Planck equation reaches equilibrium much faster than that obtained via simulation.

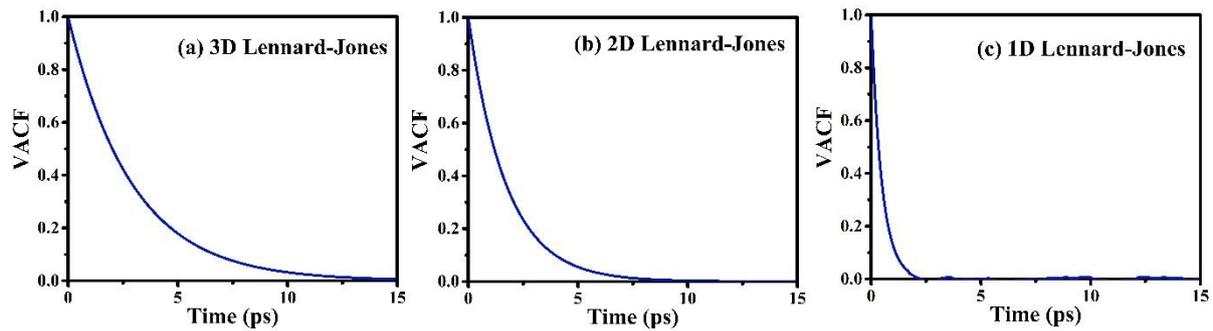

**Figure S6. The normalized velocity autocorrelation function (VACF) of (a) 3D, (b) 2D, and (c) 1D Lennard-Jones systems.**